\documentclass[aip,apl,reprint,preprintnumbers]{revtex4-1}
\usepackage{graphicx}
\usepackage{dcolumn}
\usepackage{bm}
\usepackage{color}
\usepackage{sidecap}

\begin{document}

\title{Continuous operation of high bit rate quantum key distribution}

\author{A. R. Dixon}

\altaffiliation [Also at] { Cavendish Laboratory, University of Cambridge, J. J. Thomson
Avenue, Cambridge CB3 0HE, UK.}

\author{Z. L. Yuan}
\email{zhiliang.yuan@crl.toshiba.co.uk}

\author {J. F. Dynes}

\author {A. W. Sharpe}

\author {A. J. Shields}

\affiliation{Toshiba Research Europe Ltd, Cambridge Research
Laboratory, 208 Cambridge Science Park, Milton Road, Cambridge, CB4~0GZ, UK }

\date{\today}

\begin{abstract}
We demonstrate a quantum key distribution with a secure bit rate exceeding 1 Mbit/s over 50 km fiber averaged over a continuous 36-hours period.  Continuous operation of high bit rates is achieved using feedback systems to control path length difference and polarization in the interferometer and the timing of the detection windows. High bit rates and continuous operation allows finite key size effects to be strongly reduced, achieving a key extraction efficiency of 96\% compared to keys of infinite lengths.
\end{abstract}

\pacs{03.67.Dd Quantum Cryptography; 85.60.Gz Photo detectors; 85.60.Gw Photodiodes;}

\maketitle

Quantum key distribution (QKD) promises unprecedented security for communication.\cite{bb84,ekert91,gisin02,scarani09} The security is based on physical laws rather than the supposed difficulty of computing certain mathematical problems. The BB84 protocol\cite{bb84} has been proven unconditionally secure against the most general attacks by several different approaches.\cite{shor00,renner05,koashi06,gllp} In combination with decoy pulses,\cite{hwang03,wang05,lo05} secure key distribution has been made possible using practical components, such as weak laser pulses. Moreover, recent works deal robustly with keys of finite size\cite{cai09} and also guarantee composable security,\cite{rice09} an important property if the key is to be used in an actual cryptographic implementation.

Experimentally,  distribution of secure keys has been demonstrated  at rates exceeding 1 Mbit/s between two parties separated by several kilometers.\cite{dixon08, zhang09} When combined with the one-time pad cipher this enables communication to be encrypted at this rate with ultimate security. However, it is often the case, including in these high speed reports, that key distribution is only possible for a short period of time. In this ``one shot" mode extensive set up and calibration must be performed prior to a short key exchange, after which time re-calibration is required. In addition, key rates of 1~Mbit/s are typically reached only over short fiber distances (around 10 km), and fall rapidly beyond this.\cite{dixon08,zhang09}

The first security proofs\cite{shor00,renner05,koashi06,gllp} for the BB84 protocol, and for almost all other protocols, are valid only for keys of infinite length. In practice, keys have a finite length and so parameters estimated from the key distribution have some statistical fluctuations. To guarantee security these fluctuations must be taken account of in the security proof in a pessimistic way, reducing the secure key rate compared to the infinite key case. This finite key size effect creates a dependence between key size and secure key rate; keys of greater size have smaller statistical fluctuations and so less reduction in secure rate.

The key efficiency (the secure key rate compared to the infinite key case) is a function of the key size and hence the session duration. It drops rapidly when the number of pulses sent becomes less than $10^{11}$ according to our simulation for a 50-km fiber distance. To reach a key efficiency of 95\%, the number of pulses must be greater than $7.5\times10^{11}$. This translates to a session duration of longer than one day for a megahertz-clocked QKD system.\cite{rosenberg09} Even for a high speed system,\cite{dixon08} a key generation time on the order of 10 minutes is required.  Previously, all the Mb/s systems can be operated continuously only for less than one minute due to the instability of the quantum optics. Clearly for efficient key generation it is crucial for a QKD system to operate at a gigahertz clocked rate and with long term stability.

Here, we have implemented the BB84 protocol in a high speed system built using robust commercially available components and able to operate continuously at high speed. As is becoming standard in QKD implementations based on the BB84 protocol we use weak-pulse decoy states\cite{lo05,wang05} to reduce the systems susceptibility, and thus secure key cost, to all physically possible attacks, and in particular to photon number splitting attacks.\cite{huttner95}

\begin{figure*}
\centering\includegraphics[width=1.8\columnwidth]{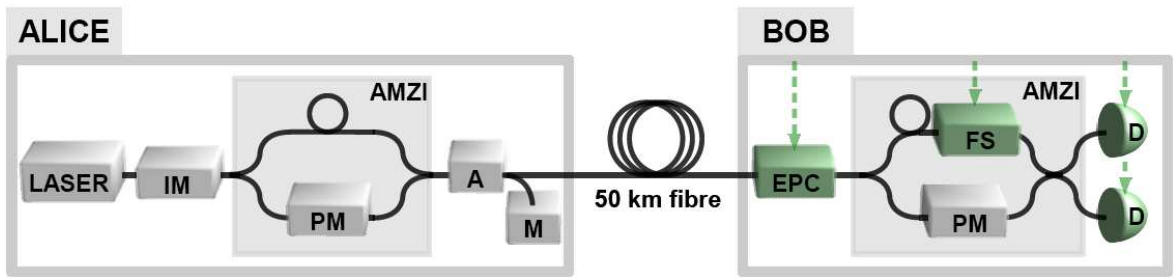}
\caption{Schematic of QKD system. IM denotes fiber intensity modulator, PM phase modulator, A attenuator, M optical power meter, AMZI asymmetric Mach-Zender interferometer, EPC electrically-driven polarization controller, FS fiber stretcher, D InGaAs APD detectors. Components in green are feedback-controlled as part of the active stabilization system.}
\label{fig:setup}
\end{figure*}

We have developed a stable GHz QKD system, outlined in Fig. 1. A fiber laser pulsed at 1 GHz produces 1550~nm photon pulses with a 15~ps width. A LiNbO$_3$ fiber intensity modulator creates the different pulse intensities required for the decoy protocol. The information is encoded on the photons phase using an asymmetrical Mach-Zender interferometer, with a phase modulator located in one arm. The total photon flux is measured using a beam splitter just before transmission into the fiber, and maintained at the correct level using a computer controlled variable attenuator. The photons are transmitted through 50 km of single-mode fiber with a measured attenuation loss of 0.2~dB/km. After transmission through the fiber the photons pass through a four-channel electronic polarization controller, to correct for any polarization drift in the fiber, and then Bob's interferometer. One arm contains a phase modulator for basis selection while the other arm contains a fiber stretcher. The fiber stretcher enables the path length to be continually altered to maintain the same path difference as in Alice's interferometer. The photons are detected by InGaAs avalanche photodiodes (APDs) gated in Geiger mode at 1~GHz and cooled to -30 $^\circ$C using a Peltier cooler, with an efficiency of 16.5\% and a dark count probability of $9\times10^{-6}$ per gate. The APDs output is processed using the self-differencing technique,\cite{yuan07} which enables much smaller avalanches to be detected and thus suppresses afterpulses, allowing for operation at high gating speeds.

A complete active stabilization scheme is implemented to enable the system to compensate for condition changes, allowing QKD sessions over arbitrarily long time durations.  This ability is crucial to minimize the effect of secure key rate reduction due to the finite key size.  Instabilities can be divided into two categories: (i) coding instability due to fiber length variations in Alice and Bob's encoder and decoder respectively, and (ii) channel transmission instability due to polarization and arrival time drifts in the quantum channel. The former adds to the quantum bit error rate (QBER), while the latter causes a drop in the bit rate. The coding instability is corrected for by using the fiber stretcher to continually alter the path difference in Bob's decoder to ensure it matches Alice's encoder.\cite{yuan05} This fiber stretcher is controlled using the QBER as a feedback signal. When the error rate is minimized the differences in interferometer path lengths are eliminated.

\begin{figure}[b]
\newpage
\centering\includegraphics[width=.8\columnwidth]{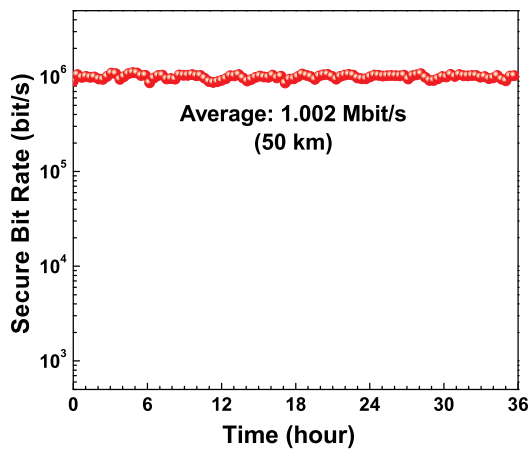}
\caption{Secure key rate as a function of time over a continuous 36 hour period for a fiber distance of 50~km.}
\label{fig:bitrate}
\end{figure}

There are two time varying parameters contributing to the transmission instability: the length and polarization characteristics of the transmission fiber linking Alice and Bob. Drift in the fiber length leads to the photon arrival time at the detectors moving outside of the active time window, reducing the bit rate. Drift in the polarization leads to photons traveling through non-interfering paths, either through both long arms or both short arms in Alice and Bob's interferometers, and thus reducing the bit rate. To compensate for these drifts, the detector count rates are used as a feedback signal to adjust the delay position of the detector gate as well as the polarization controller state.  The active stabilization operates continuously alongside key distribution, and as such there is no duty cycle or key rate reduction due to its use.

The decoy BB84 protocol\cite{wang05,lo05} with three different pulse intensities is implemented. Signal pulses ($\mu = 0.5$ photons per pulse, sent with 98.83\% probability) are used for distributing the key while two different decoy pulses ($\nu_1 = 0.1$ and $\nu_2=0.0007$ photons per pulse sent with 0.78\% and 0.39\% probability respectively) are used only to characterize the quantum channel, to prevent photon number splitting attacks. The photon fluxes and transmission probabilities were found using numerical simulations to optimize the secure key rate. We implement a recently developed finite key version\cite{rice09} of Koashi's security proof\cite{koashi06} based on the uncertainty principle. The keys generated using this proof have universally composable security,\cite{muller09} with each key having a chosen failure probability of $\epsilon$, and it also does not assume an underlying Gaussian distribution for statistics, instead utilizing the full binomial distribution.\cite{clopper34} Here, we choose $\epsilon=10^{-7}$.

Figure 2 shows the secure key distribution rate as a function of time, obtained over a period of 36 hours continuous operation. A secure key is distilled every 20 minutes, a time duration that is sufficiently long to achieve 96\% of the theoretical key rate for an infinitely long QKD session. A total key size of $1.3\times10^{11}$ bits is distributed at an average secure key rate of 1.002 Mbit/s. This rate is a factor of two to three orders of magnitude improvement as compared to the stable SECOQC network,\cite{peev09} and a factor of four higher than the previous record obtained with a non-stabilized system.\cite{dixon08}

\begin{figure}[t]
\centering\includegraphics[width=.8\columnwidth]{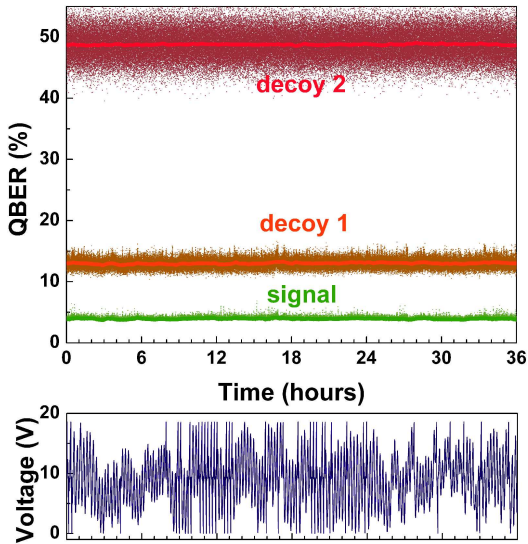}
\caption{(Top) Quantum bit error rate (QBER) as a function of time for signal and decoy pulses separately. The instantaneous value calculated every second is shown as points, with the aggregate value used for key distillation every 20~mins shown as the solid lines. (Bottom) Voltage of the fiber stretcher, located in Bob's interferometer, over the same period.}
\label{fig:qber}
\end{figure}

Figure 3 shows the QBERs, including both instantaneous and average values. The instantaneous  values were determined per second and used to stabilize the interferometer path lengths, while the average QBER indicates the value aggregated over 20 minutes and is used for distilling the secure key. While the effects of statistical fluctuations can be clearly seen on the instantaneous QBER of the decoy states (the count rate of the second decoy state is only a few hundred counts per second), the stability of the rate is remarkable, with no spikes outside of expected statistical fluctuations evident. The mean signal error rate is 3.85\%, with 90\% of values within 10\% of this, and no value measured above 6.86\% for the entire 36 hour period.

\begin{figure}[t]
\centering\includegraphics[width=.8\columnwidth]{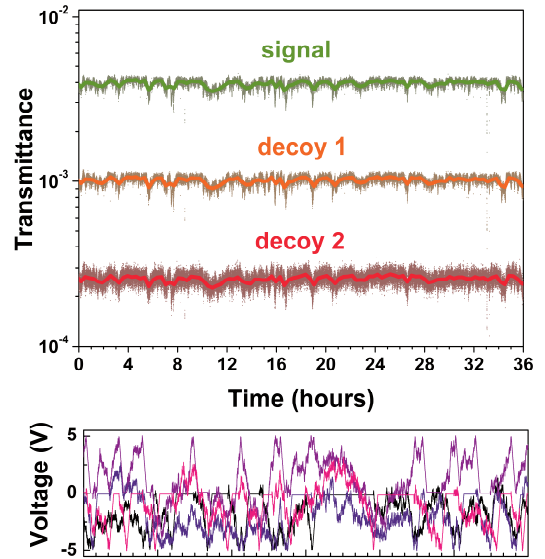}
\caption{(Top) Transmittance (detection probability per gate) as a function of time for signal and decoy pulses separately. The instantaneous value calculated every second is shown as points, with the aggregate value used for key distillation every 20~mins shown as the solid lines. (Bottom) Voltage of each of the 4 channels of Bob's polarisation controller over the same period.}
\label{fig:transmittance}
\end{figure}

Figure 4 shows the transmittance (detection probability per pulse sent) of the signal and decoy states. Both the per second rates (points) and aggregate rates (lines) are stable over both the short and long term; without active stabilization the transmittances slowly decrease over tens of minutes (not shown) due to polarization drift in the fiber, before falling to near zero after a few hours due to the photon time of arrival shifting outside of the detection gate.

Finally, we comment on the source requirement for decoy protocols. The first prerequisite is the indistinguishability of signal and decoy pulses in both spectral and time domains. This is ensured by the use of a single laser with intensity modulation. The second prerequisite is the stability of the light source intensities.\cite{yuan07,wang09} In this experiment, the total intensity transmitted is controlled by a feedback loop. The ratio of decoy to signal count rate is stable with a standard deviation of less than 0.5\% among all QKD sessions, suggesting a precise control of source intensities.

In conclusion, we have demonstrated a QKD system able to operate continuously and autonomously over long periods of time, at a record key distribution rate of 1 Mbit/s over 50 km of fiber. We believe the stable and high speed implementation, achieved using only commercially available and robust devices, in particular InGaAs APDs, is highly suitable for practical implementations.

The authors thank V. Scarani for useful discussions on QKD with finite resources.




\begin{thebibliography} {99}

\bibitem{bb84} C. H. Bennett and G. Brassard, \textit{Proceedings of the IEEE International Conference on Computers, Systems and Signal Processing,} Bangalore, India, pp. 175-179 (1984).

\bibitem{ekert91} A. K. Ekert, Phys. Rev. Lett. \textbf{67}, 661 (1991).

\bibitem{gisin02} N. Gisin, G. Ribordy, W. Tittel adn H. Zbinden, Rev. Mod. Phys. \textbf{74}, 145 (2002).

\bibitem{scarani09} V. Scarani, H. Bechmann-Pasquinucci, N. J. Cerf, M. Du\v{s}ek, N. L\"{u}tkenhaus and M. Peev, Rev. Mod. Phys. \textbf{81}, 1301 (2009).

\bibitem{shor00} P. W. Shor and J. Preskill, {Phys. Rev. Lett.} \textbf{85}, 441 (2000).

\bibitem{renner05} R. Renner, N. Gisin and B. Kraus, {Phys. Rev. A} \textbf{72}, 012332 (2005).

\bibitem{koashi06} M. Koashi, {arXiv:quant-ph}/0609180v1 (2006).

\bibitem{gllp} D. Gottesman, H. K. Lo, N. L\"utkenhaus, and J. Preskill, {Quantum Inf. Comput.} \textbf{4}, 325 (2004).

\bibitem{hwang03} W. Y. Hwang, {Phys. Rev. Lett.} \textbf{91}, 057901 (2003).

\bibitem{wang05} X. B. Wang, {Phys. Rev. Lett.} \textbf{94}, 230503 (2005).

\bibitem{lo05} H. K. Lo, X. Ma, and K. Chen, {Phys. Rev. Lett.} \textbf{94}, 230504 (2005).

\bibitem{cai09} R. Y. Q. Cai and V. Scarani, {New J. Phys.} \textbf{11}, 045024 (2009).

\bibitem{rice09} P. Rice and J. Harrington, {arXiv}:/0901.0013v2 (2009).

\bibitem{dixon08} A. R. Dixon, Z. L. Yuan, J. F. Dynes, A. W. Sharpe and A. J. Shields, {Opt. Express} \textbf{16}, 18790 (2008).

\bibitem{zhang09} Q. Zhang, H. Takesue, T. Honjo, K. Wen, T. Hirohata, M. Suyama, Y. Takiguchi, H. Kamada, Y. Tokura, O. Tadanaga, Y. Nishida, M. Asobe and Y Yamamoto, {New J. Phys.} \textbf{11}, 045010 (2009).

\bibitem{rosenberg09} D. Rosenberg,  C. G. Peterson, J. W. Harrington, P. R. Rice, N. Dallmann, K. T. Tyagi, K. P. McCabe, S. Nam, B. Baek, R. H. Hadfield, R. J. Hughes and J. E. Nordholt, {New J. Phys.} \textbf{11}, 045009 (2009).

\bibitem{huttner95} B. Huttner, N. Imoto, N. Gisin and T. Mor, {Phys. Rev. A} \textbf{51}, 1863 (1995).

\bibitem {yuan07} Z. L. Yuan, B. E. Kardynal, A. W. Sharpe, and A. J. Shields, Appl. Phys. Lett. \textbf{91}, 041114 (2007).

\bibitem{yuan05} Z. L. Yuan and A. J. Shields, {Opt. Express} \textbf{13}, 660 (2005).

\bibitem{muller09} J. M\"uller-Quade and R. Renner, {New J. Phys.} \textbf{11}, 085006 (2009).

\bibitem{clopper34} C. Clopper and E. S. Pearson, {Biometrika} \textbf{26}, 404 (1934).


\bibitem{peev09} M. Peev, C. Pacher, R. Alleaume, C. Barreiro, W. Boxleitner, J. Bouda, R. Tualle-Brouri, E. Diamanti, M. Dianati, T. Debuisschert, J. F. Dynes, S. Fasel, S. Fossier, M. Fuerst, J.-D. Gautier, O. Gay, N. Gisin, P. Grangier, A. Happe, Y. Hasani, M. Hentchel, H. H\"ubel, G. Humer, T. L\"anger, M. Legre, R. Lieger, J. Lodewyck, T. Lorünser, N. L\"utkenhaus, A. Marhold, T. Matyus, O. Maurhart, L. Monat, S. Nauerth, J.-B. Page, E. Querasser, G. Ribordy, A. Poppe, L. Salvail, S. Robyr, M. Suda, A. W. Sharpe, A. J. Shields, D. Stucki, C. Tamas, T. Themel, R. T. Thew, Y. Thoma, A. Treiber, P. Trinkler, F. Vannel, N. Walenta, H. Weier, H. Weinfurter, I. Wimberger, Z. L. Yuan, H. Zbinden, and A. Zeilinger, {New J. Phys.} \textbf{11}, 075001 (2009).

\bibitem{yuan07} Z. L. Yuan, A. W. Sharpe and A. J. Sheilds, Appl. Phys. Lett. \textbf{90}, 011118 (2007).

\bibitem{wang09} X. B. Wang, L. Yang, C. Z. Peng and J. W. Pan, New. J. Phys. \textbf{11}, 075006 (2009)

\end{thebibliography}
\end{document}